\documentclass[%
 reprint,
 amsmath,amssymb,
 aps,
]{revtex4-2}

\usepackage{graphicx}
\usepackage{dcolumn}
\usepackage{bm}
\usepackage{ulem}
\usepackage[switch]{lineno}
\usepackage[margins]{trackchanges}
\usepackage{siunitx}
\begin{document}

\preprint{APS/123-QED}

\newcommand{\hs}{\hspace{6pt}}
\newcommand{\eps}{\epsilon}
\newcommand{\beps}{\boldsymbol{\epsilon}}
\newcommand{\bdel}{\boldsymbol{\delta}}
\newcommand{\dd}{\partial}
\newcommand{\q}{\theta}
\newcommand{\DD}{\mathcal{D}}
\newcommand{\uu}{\mathbf{u}}
\newcommand{\UU}{\mathbf{U}}
\newcommand{\VV}{\mathbf{V}}
\newcommand{\WW}{\mathbf{W}}
\newcommand{\MM}{\mathbf{M}} 
\newcommand{\hh}{\mathbf{h}} 
\newcommand{\ZZ}{\mathbf{Z}}
\newcommand{\SSS}{\mathbf{S}}
\newcommand{\QQ}{\mathbf{Q}}

\newcommand{\XX}{\vec{X}} 
\newcommand{\EE}{\mathbf{E}}
\newcommand{\YY}{\mathbf{Y}}

\newcommand{\NN}{\vec{N}}

\newcommand{\HH}{\vec{H}}
\newcommand{\CL}{\hat{\boldsymbol{\mathcal{L}}}}
\newcommand{\CLd}{\hat{\boldsymbol{\mathcal{L}}}^{\dagger}}

\newcommand{\tens}[1]{\bar{\bar{#1}}}

\newcommand{\tuu}{\tilde{\mathbf{u}}}
\newcommand{\TUU}{\tilde{\mathbf{U}}}
\newcommand{\OO} {\mathcal {O}}
\newcommand{\ve}[1]{\vec{e#1}}
\newcommand{\sig}{\sigma}
\newcommand{\la}{\lambda}

\newcommand{\rr}{\rho}
\newcommand{\HP}{\hat{\mathbf{P}}}
\newcommand{\HQ}{\hat{\mathbf{Q}}}
\newcommand{\HL}{\hat{\mathbf{L}}}
\newcommand{\HPI}{\hat{\mathbf{\Pi}}}
\newcommand{\bI}{\mathbf{I}}
\newcommand{\bF}{\mathbf{F}}
\newcommand{\GG}{\mathcal{G}}
\newcommand{\ls}{\lambda_\s}
\newcommand{\RR}{\mathcal{R}}
\newcommand{\RRR}{\mathbf{R}}
\newcommand{\TRK}{\mathrm{Tr} \mathbf{K}}
\newcommand{\bo}{\otimes}
\newcommand{\dg}{|g|}
\newcommand{\vx}{\vec{x}}
\newcommand{\hyp}[1]{(\textbf{#1})}
\newcommand{\nn}{\nonumber}

\newcommand{\bR}{\SSS}
\newcommand{\bP}{\mathbf{P}}
\newcommand{\tP}{\tilde{\mathbf{P}}}
\newcommand{\bpsi}{\boldsymbol{\psi}}
\newcommand{\bs}{\boldsymbol{\sigma}}
\newcommand{\bone}{\boldsymbol{1}} 
\newcommand{\bY}{\mathbf{Y}}
\newcommand{\BLA}{\boldsymbol{\Lambda}}

\newcommand{\braket}[2]{\left\langle #1 \mid #2 \right\rangle  }
\newcommand{\bra}[1]{\left\langle #1 \right\rvert  }
\newcommand{\ket}[1]{\left\lvert #1 \right\rangle  }

\newcommand{\bbraket}[2]{\langle \langle#1 \mid #2 \rangle \rangle }
\newcommand{\bbra}[1]{\langle\langle #1 \mid  }
\newcommand{\kket}[1]{\mid #1 \rangle\rangle  }

\newcommand{\bsub}{\begin{subequations}}
\newcommand{\esub}{\end{subequations}}
\newcommand{\ba} {\begin{align} }
\newcommand{\ea}{ \end{align} }

\newcommand{\zeroket}{| \boldsymbol{0} \rangle}

\newcommand{\etal}{\textit{et al.}}

\newcommand{\hmu}{{\hat{\mu}}}
\newcommand{\hnu}{{\hat{\nu}}}

\newcommand{\hA}{{\hat{A}}}
\newcommand{\hB}{{\hat{B}}}
\newcommand{\ha}{{\hat{a}}}
\newcommand{\hb}{{\hat{b}}}

\newcommand{\CP}{\mathcal{P}}
\newcommand{\CE}{\mathcal{E}}
\newcommand{\CM}{\mathcal{M}}
\newcommand{\CX}{\mathcal{X}}

\newcommand{\hpi}{ \hat{\pi}}

\newcommand{\omf}{ \zeta_0}
\newcommand{\CC}{\mathbf{C}}
\newcommand{\BTH}{\boldsymbol{\Theta}}
\newcommand{\XI}{\boldsymbol{\Xi}}
\newcommand{\bom}{ \boldsymbol{\omega}}
\newcommand{\bka}{ \boldsymbol{\kappa}}
\newcommand{\bxi}{ \boldsymbol{\xi}}
\newcommand{\KK}{\mathbf{K}}
\newcommand{\BB}{\mathbf{B}}
\newcommand{\TT}{\mathbf{T}}
\newcommand{\vv}{\mathbf{v}}
\newcommand{\xx}{\mathbf{x}}
\newcommand{\ee}{\mathbf{e}}
\newcommand{\ii}{\mathbf{i}}
\newcommand{\txx}{\tilde{\mathbf{x}}}
\newcommand{\TXX}{\tilde{\mathbf{X}}}

\newcommand{\HGamma}{\hat{ \Gamma}}

\newcommand{\fwd}[1]{ \bar{#1} }
\newcommand{\cen}[1]{\langle{ #1} \rangle}

\renewcommand{\aa}{\alpha}
\newcommand{\bb}{\beta}
\newcommand{\cc}{\gamma}

\newcommand{\Tr}{\mathrm{Tr}}

\newcommand{\red}[1]{ {\color{red} #1 }}
\newcommand{\orange}[1]{ {\color{orange} #1 }}
\newcommand{\green}[1]{ {\color{ForestGreen} #1 }}

\newcommand{\reals}{\mathbb{R}}

\title{Intrinsic memory limits resonant control of complex cardiac spiral waves}
\author{Navneet Roshan$^{1}$}

\author{Hans Dierckx$^{2,3,4}$}

\author{Rupamanjari Majumder$^{1,*}$}
\affiliation{
 $^1$Nantes Université, INSERM, l’institut du thorax, F-44000 Nantes, France.
}
\affiliation{$^2$Laboratory of Experimental Cardiology, Leiden University Medical Center, the Netherlands 
}
\affiliation{$^3$Mathematical Institute, Leiden University, the Netherlands 
}
\affiliation{$^4$ Institute of Physics (LION) Leiden University, the Netherlands 
}
\email{rupamanjari.majumder@univ-nantes.fr}
\date{\today}

\begin{abstract}

Resonant forcing is known to induce controlled drift of phase-locked spiral waves in excitable media. Here, we show that intrinsic memory fundamentally alters this paradigm in complex meandering spirals typical of cardiac tissue, where phase-dependent responses to successive perturbations over the meander cycle interfere destructively to produce metastable quasi-stationarity or constructively to induce stable confinement. These dynamical insights motivate an improved low-energy multi-pulse feedback protocol that robustly drives spiral waves to termination through sustained directed drift.
\end{abstract}

\maketitle
%\linenumbers
 
Spiral waves are self-organized rotating patterns observed across a wide range of physical, chemical, and biological systems~\cite{Zaikin1970,Winfree1972,pertsov1993spiral,Gerisch1987,Kastberger2008}. Their emergence is a hallmark of nonlinear excitable media. Many such media exhibit intrinsic memory, i.e., recovery and restitution couple present excitability to prior activation history~\cite{Cherry08,toal2009short,jordan2004determining,weiss2006pulsus}. Intrinsic memory influences wave propagation, stability, and susceptibility to perturbations~\cite{decker2009properties}, often rendering spiral waves drivers of complex dynamics. In the heart, spiral waves underlie lethal arrhythmias, with tissue-specific restitution properties often governing transitions between regular and chaotic behavior. Their sensitivity to local perturbations makes control inherently challenging, as interventions often promote destabilization instead of suppression of these waves.

Resonance-based feedback control (RFC) is a well-established strategy for controlling spiral waves in excitable media, using weak, phase-synchronized perturbations to steer spiral-wave dynamics~\cite{Davydov1988,Mikhailov1994,Biktashev:1995, Biktashev:1995b}. 

The theoretical framework for RFC has been extended beyond rigidly rotating spirals to periodically forced meandering spirals, where weak perturbations can be described in terms of response functions~\cite{Mantel1996,Biktasheva2010}. RFC can organize such spirals into resonance attractors, with sustained drift along stable orbits~\cite{Zykov:1997,Grill1995} governed by feedback parameters and domain geometry~\cite{Braune2000,Zykov2002,Zykov2004}. In chemical excitable media and simplified reaction--diffusion systems, successive perturbations remain sufficiently coherent to sustain organized drift, although at velocities substantially below the wave-propagation speed~\cite{Braune2000,Zykov2004}. These frameworks, however, do not account for pronounced intrinsic memory, where history-dependent responses may cause successive resonant stimuli to reinforce or oppose one another rather than accumulate into coherent translation. Whether organized resonant drift persists under such conditions therefore remains unclear. This raises a central nonlinear-dynamics question: can weak resonant forcing sustain directed control when intrinsic memory shapes the response to successive perturbations?

Here, we show that intrinsic memory disrupts the coherent accumulation underlying conventional resonant drift. Although slow spiral-wave motion persists, it acquires a fundamentally different dynamical origin: phase-dependent responses to successive stimuli nearly cancel over the meander cycle, producing a metastable quasi-stationary state rather than sustained drift along a resonance attractor. This state arises from the intrinsic temporal dynamics and persists across domain sizes, heterogeneous media, and realistic cardiac geometry. Even during orbital precession, intermittent emergence of quasi-stationarity can disrupt the orbit and prevent sustained coherent drift. Conversely, constructive accumulation of phase-dependent responses produces sensor-centered confinement. Thus, intrinsic memory transforms resonant forcing from coherent drift into phase-structured interference, fundamentally reorganizing spiral-wave dynamics and limiting conventional RFC. Building on this mechanism, we develop a strategy that substantially improves spiral-wave control.

To study resonant feedback pacing in meandering spiral waves whose dynamics is restitution-driven, we used the three-variable Fenton–Karma model (3vFK) \cite{fenton1998vortex}. This model reproduces spiral waves with core dynamics similar to detailed ionic models such as the Luo–Rudy model \cite{luo1991model} for guinea pig heart and Courtemanche model \cite{Courtemanche1998} for human atria. The governing equations are

$\partial_{t} u = \nabla \cdot(D_0 \nabla u)-\mathcal{J_{\text{ion}}}$, 
$\partial_{t} v = \Theta\left(u_{c}-u\right)(1-v) / \tau_{v}^{-}(u)-\Theta\left(u-u_{c}\right) v / \tau_{v}^{+}$,  
$\partial_{t} w = \Theta\left(u_{c}-u\right)(1-w) / \tau_{w}^{-}-\Theta\left(u-u_{c}\right) w / \tau_{w}^{+}$,

where $u\in[0,1]$ is the dimensionless transmembrane voltage, dimensional voltage ($V$) was obtained via \(V = 100u - 85\) (mV), $v$ and $w$ are fast and slow gating variables, and $\mathcal{J}{\text{ion}}=J_{\mathrm{fi}}+J_{\mathrm{so}}+J_{\mathrm{si}}$. Diffusion constant $D_0=0.001$~\si{cm^2/ms} \cite{shajahan2007spiral}. Parameter values yield an action potential duration of $\sim210$~ms (90\% repolarization) and wavelength $13.7$~cm at a pacing cycle length of $1000$~ms, consistent with mammalian ventricular tissue \cite{ten2004model}. Unless stated otherwise, simulations were performed on a $8\times8$~\si{cm^2} domain using finite difference method with spatial resolution $0.02$~cm and time step $\delta t=0.05$~ms.

A defining dynamical feature of multilobed meander is the variation in wavefront–waveback interactions near the spiral tip, accompanied by phase-dependent changes in propagation speed along the floral meander trajectory. In our case, the spiral tip followed a hybrid trajectory combining linear motion and sharp rotational turns at petal tips. The full 5-petal meander pattern was covered in $T_0 = 384$ \si{ms}. To implement RFC, we continuously monitored the excitation at a single point (sensor) and applied a uniform global square-pulse current stimulus ($I_{\mathrm{stim}}$) whenever the local voltage crossed a prescribed threshold, $u_{\mathrm{crit}}=0.2$, with $\dot{u}\geq0.015$.
 During one $T_0$-cycle our spiral completed approximately $3$ rotations, thereby exciting the sensor $3$ times. Between successive rotations, the tip accelerated and decelerated, producing oscillatory velocity. These variations were governed by changes in the diastolic interval, which link the spiral’s motion to intrinsic dynamical memory.
\begin{figure}
\centering
	\includegraphics[width=8.54cm]{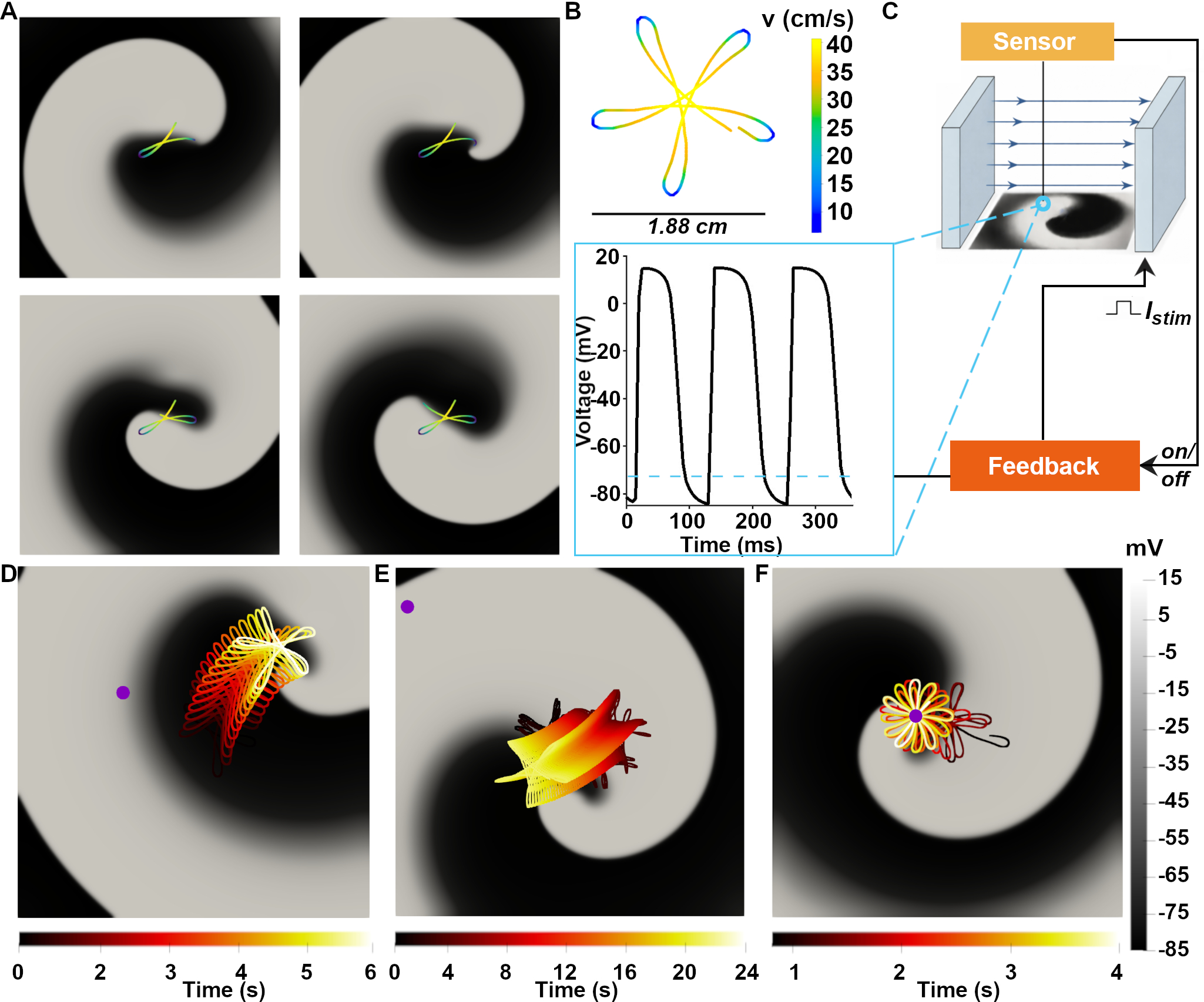}
	\caption{Meander-driven wavefront–waveback interaction and feedback dynamics.
(A) Wavefront–waveback proximity during a single petal traversal.
(B) Tip speed: maximal near the core, decreasing along petals, minimal at tips.
(C) RFC protocol: a local sensor triggers uniform stimulation when $u$ crosses threshold ($u_{\textrm{crit}}=0.2$, $\dot{u}\geq0.015$); three activations per cycle.
(D–F) Tip trajectories for different sensor positions (purple dots), showing three distinct dynamical regimes.
}
	\label{fig:Fig1}
\end{figure}
\begin{figure*}
\centering
	\includegraphics[width=14cm]{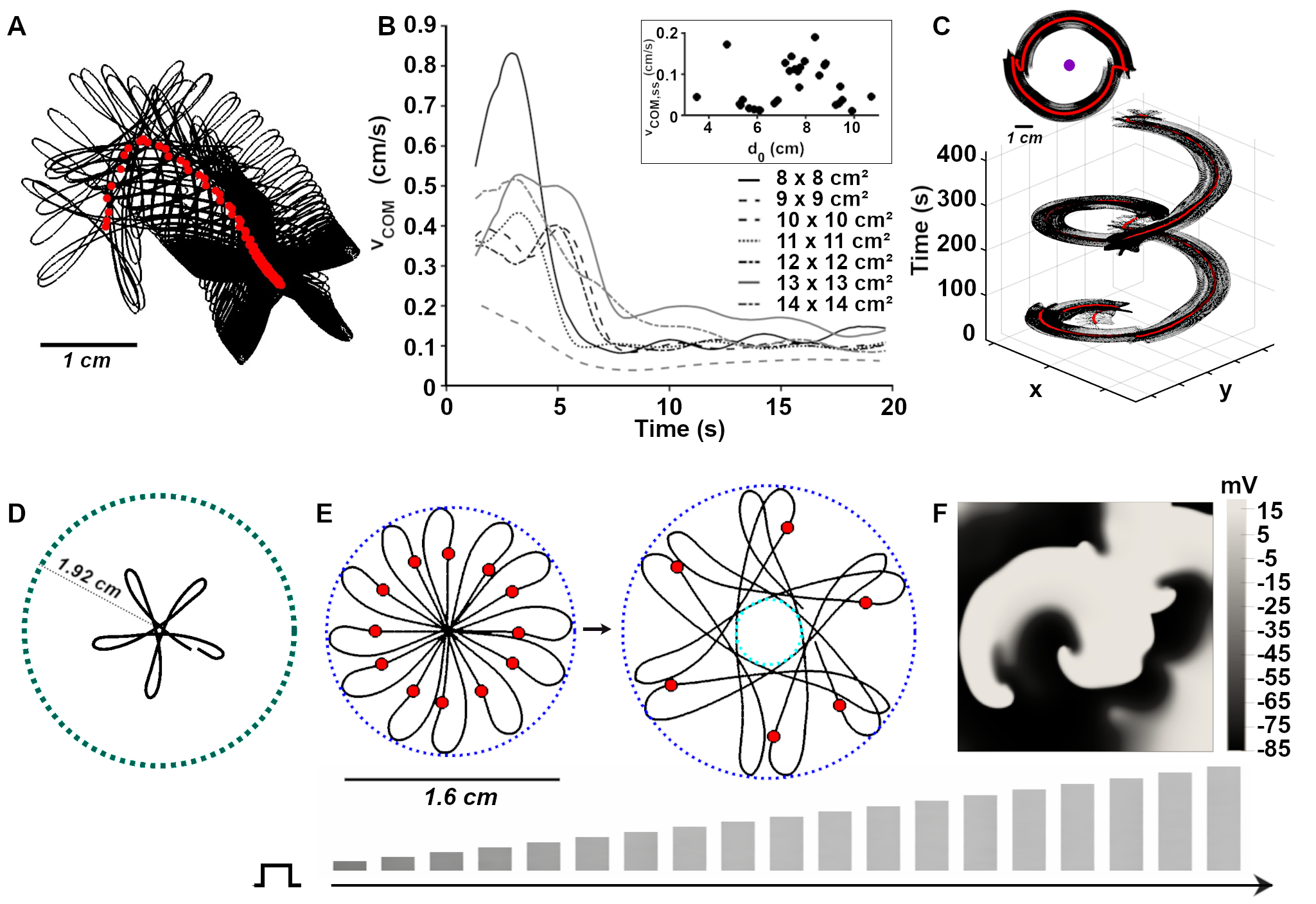}
	\caption{Metastable quasi-stationary dynamics and confinement under RFC.
(A) Spiral trajectory in an $8\times8~\mathrm{cm^2}$ homogeneous domain showing transition from resonant drift to a metastable quasi-stationary state. Filled circles denote the cycle-averaged spiral COM, with increasing density indicating slowing.
(B) Time evolution of COM speed $v_{\mathrm{COM}}$, collapsing to near-stationary values across domain sizes and sensor locations. Inset: steady-state speed $v_{\mathrm{COM,ss}}$ vs sensor distance $d_0$.
(C) Metastability in large domains, with intermittent excursions into hypermeander or drift; inset: distorted trajectory from regime switching.
(D) Basin of attraction for confinement with the sensor near the spiral core.
(E) Tip trajectories under confinement for increasing stimulus amplitude, showing progressive opening of the floral pattern and growth of inner and outer core radii, red dots indicate the phases at which sensor was triggered.
(F) Confinement uniquely induces wave breakup at amplitudes subcritical in the quasi-stationary and drift regimes.}
	\label{fig:Fig2}
\end{figure*}
Figure~\ref{fig:Fig1}A shows wavefront–waveback proximity during a single petal traversal. The spiral tip speed (Fig.~\ref{fig:Fig1}B) peaks near the flower center, decreases along the petals, and is minimal at the petal tips. Under RFC (Fig.~\ref{fig:Fig1}C), the response depends sensitively on sensor location (Figs.~\ref{fig:Fig1}D–F) and organizes into three distinct dynamical regimes. 
These regimes show markedly different drift characteristics, revealing a hierarchy of responses that departs from classical resonant-drift predictions. Consequently, the behavior of complex meandering cardiac spirals cannot be inferred from rigid circular-core spirals, where a well-separated excitable gap suppresses cumulative memory and perturbation effects.

Our studies revealed that when present, resonant drift was either confined to relatively small simulation domains (approximately 2.5–3 times the unperturbed core diameter) or gives way to a quasi-stationary state characterized by ultra-slow drift. To quantify these dynamics, we investigated the motion of the spiral center of mass (COM), which we defined as a cycle-averaged position of the spiral tip over successive intervals of duration $T_0 = 384$ \si{ms}. Figure~\ref{fig:Fig2}A illustrates the rapid transition of the spiral from normal drift motion to near stationarity in an \(8\times 8~\mathrm{cm^2}\) homogeneous domain. Filled red circles denote the spiral COM, and the progressive densification of the trajectory, evidenced by clustering of the COM markers, marks the onset of the quasi-stationary regime.

\begin{figure*}[t]
	\centering
	\includegraphics[width=15cm]{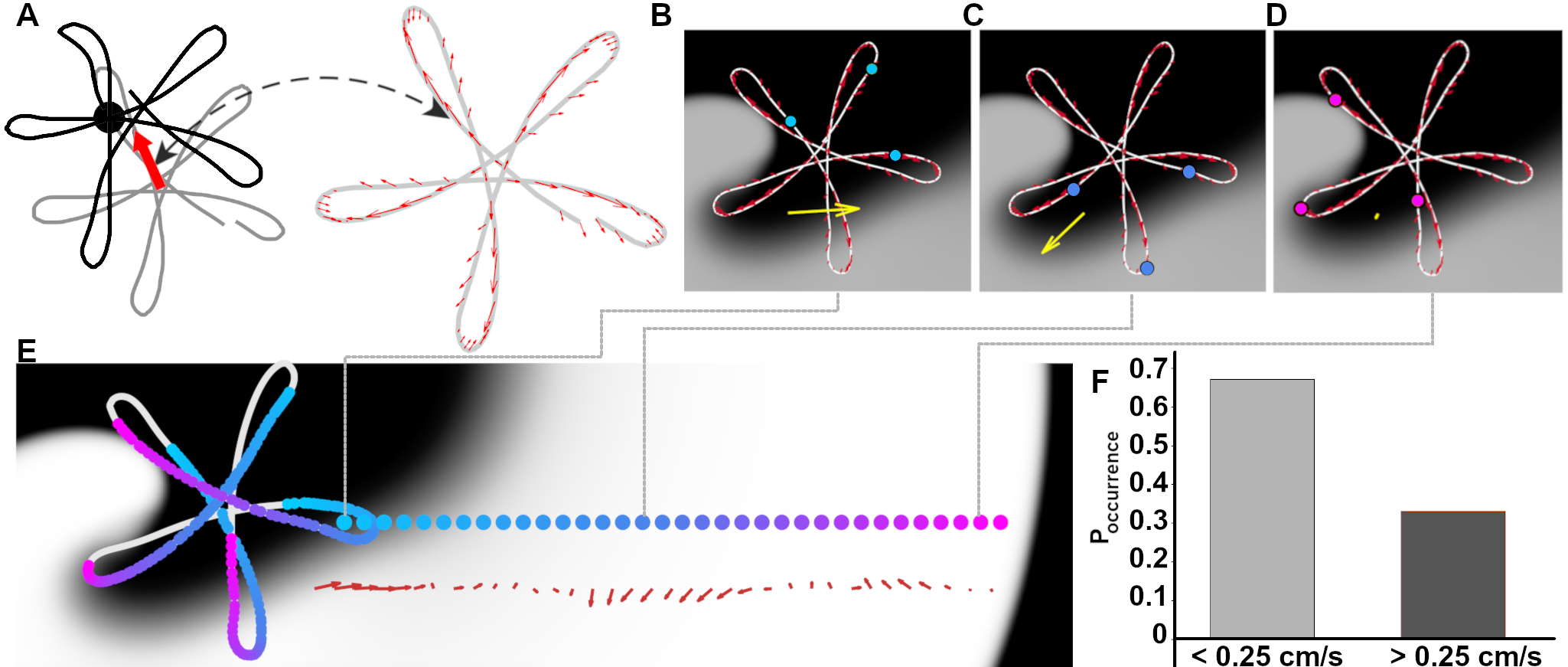}
	\caption{Phase-structured impulse responses govern attraction, repulsion, and quasi-stationary dynamics under RFC.
(A) Phase-dependent COM displacements induced by a single uniform pulse along the five-petal meander; arrows denote shift vectors.
(B,C) Net attraction (B) or repulsion (C) from constructive summation of phase-specific shifts, depending on sensor position.
(D) Quasi-stationary response from near-cancellation of phase-dependent shifts, yielding ultra-slow motion.
(E) Geometric interpretation: three stimulated phases per cycle and resulting net COM displacement (red arrows) from vector summation.
(F) Probability distribution of dynamical regimes from instantaneous COM velocity, showing dominance of the quasi-stationary state ($v_{\mathrm{COM}}<0.25$~cm/s) over sustained drift.
}
	\label{fig:Fig3} 
\end{figure*}
We found that, across domain sizes and sensor locations, the COM speed $v_{\mathrm{COM}}$ collapses to near-stationary values (Fig.~\ref{fig:Fig2}B), with the steady-state velocity $v_{\mathrm{COM,ss}}$ only weakly dependent on sensor distance $d_0$ (inset). This robust quasi-stationary regime poses a major obstacle to boundary-mediated termination. In larger domains, it manifests as ultra-slow orbital precession, consistent with Ref.~\cite{Braune2000}. However, in systems with pronounced intrinsic memory, such as cardiac tissue, intermittent transitions into drift or hypermeander disrupt the orbit before the spiral returns to ultra-slow motion, revealing the metastable nature of quasi-stationarity and a form of orbital dynamics not reported previously (Fig.~\ref{fig:Fig2}C).

A distinct response arises when the sensor lies within a finite basin around the spiral core (Fig.~\ref{fig:Fig2}D). Here, feedback induces confinement rather than drift: the spiral tip converges on and latches to the sensor, rendering the COM immobile. Within this regime, the trajectory is strongly deformed; increasing stimulus amplitude opens the floral pattern and expands core radii (Fig.~\ref{fig:Fig2}E), eventually triggering wave breakup at subthreshold amplitudes (Fig.~\ref{fig:Fig2}F). Together, these results show that RFC reorganizes spiral-wave dynamics into qualitatively distinct regimes governed primarily by the position of the spiral core relative to the sensor. Importantly, both quasi-stationarity and confinement persist under ionic and conduction heterogeneity and in complex cardiac geometry (End Matter, Fig.~\ref{fig:Supp1}), showing that these behaviors are dynamical consequences of RFC rather than features of idealized homogeneous domains.

To uncover the mechanism underlying these emergent behaviors, we quantified the COM displacement induced by phase-targeted perturbations and constructed the corresponding shift map. We discretized the five-petal meander cycle (384~ms; Fig.~\ref{fig:Fig1}B) into 100 phase bins. At each phase, we applied a single, global square pulse (amplitude $18$~\si{nA/\micro F}, duration $0.5$~ms), and the resulting COM displacement defined the shift vector. Repeating over all phases yielded the full shift diagram for the five-petal meander (Fig.~\ref{fig:Fig3}A). Note that the resulting shift vectors vary in direction and magnitude, and their accumulation under feedback pacing yields three outcomes: net attraction (Fig.~\ref{fig:Fig3}B), repulsion (Fig.~\ref{fig:Fig3}C), or near-cancellation with minimal motion (Fig.~\ref{fig:Fig3}D). These behaviors arise from vector summation over the three stimulated phases per cycle (Fig.~\ref{fig:Fig3}E), producing an inward bias within the attraction basin, an outward bias at larger distances, and near-cancellation in between. The resulting quasi-stationary band separates attraction and repulsion, with sustained drift becoming rare and transient (Fig.~\ref{fig:Fig3}F). Importantly, this suppression of net drift does not simply reflect the absence of resonant stimulation. Even when feedback repeatedly samples well-defined phases of the meander cycle, the corresponding phase-dependent displacement vectors need not align and can nearly cancel over a cycle, suppressing net translation. Intermittent departures from this phase organization account for the metastable disruption of orbital motion. 

The net feedback over one $T_0$ cycle gives rise to an iterated map for the distance $R$ of the COM relative to the sensor and the orientation (rotation phase) of the meander flower $\varphi$:
\begin{subequations} \label{eq:Rphi}
\begin{align}
R_{n+3} &= R_n + I_{stim} f(R_n, \varphi_n) \label{eq:R}\\
\varphi_{n+3} &= \varphi_n + \delta +  I_{stim} g(R_n, \varphi_n)  \label{eq:phi}
\end{align}
\end{subequations}
Here, $\delta$ is the angular deficit, i.e. the difference in radians between a full cycle and the angular distance spanned by a near-complete flower (5 petals); it leads to a slow precession of the flower at angular frequency $\delta / T_0$.  
Without phase-locking, one can average over the precession period, leading to $R_{n+3} = R_n + I_{stim} \bar{f}(R_n)$, implying monotonous drift away or towards the center, but also orbital motion at radius $R_*$ if $\bar{f}(R_*) = 0$, as documented before \cite{Grill1995} without a mathematical derivation. For sufficiently large $I_{stim}$, Eq. \eqref{eq:phi} additionally allows phase-locking \cite{Grill1995, Dierckx:2017PRL}. If $\bar{f}(R_n) < 0$ for small $R$, then the sensor can trap the spiral as in Fig.~\ref{fig:Fig1}F. Intermittent drift in Fig.~\ref{fig:Fig2}C arises when there is an almost-stable radius $R_*$ for which $f(R_*, \varphi_n) \approx 0$ but changes sign depending on $\varphi$. Then, the slow precession causes a state staying long at $R_*$, but then suddenly deviating from this value, similar to intermittent trapping observed with heterogeneities \cite{Rappel2022}.  

Thus, RFC reorganizes spiral-wave dynamics through phase-dependent interference arising from intrinsic memory and restitution.

Our findings suggest two key requirements for effective RFC. First, increasing the stimulus amplitude is not a viable strategy for accelerating drift, as sufficiently strong subthreshold perturbations can trigger wave breakup (Fig.~\ref{fig:Fig2}E,F). Second, the shift vectors form alternating spatial bands of fast and slow motion as the sensor--core distance varies (Fig.~\ref{fig:Fig3}E), allowing the spiral to become trapped in quasi-stationary regions. 
\begin{figure}
	\includegraphics[width=8.54cm]{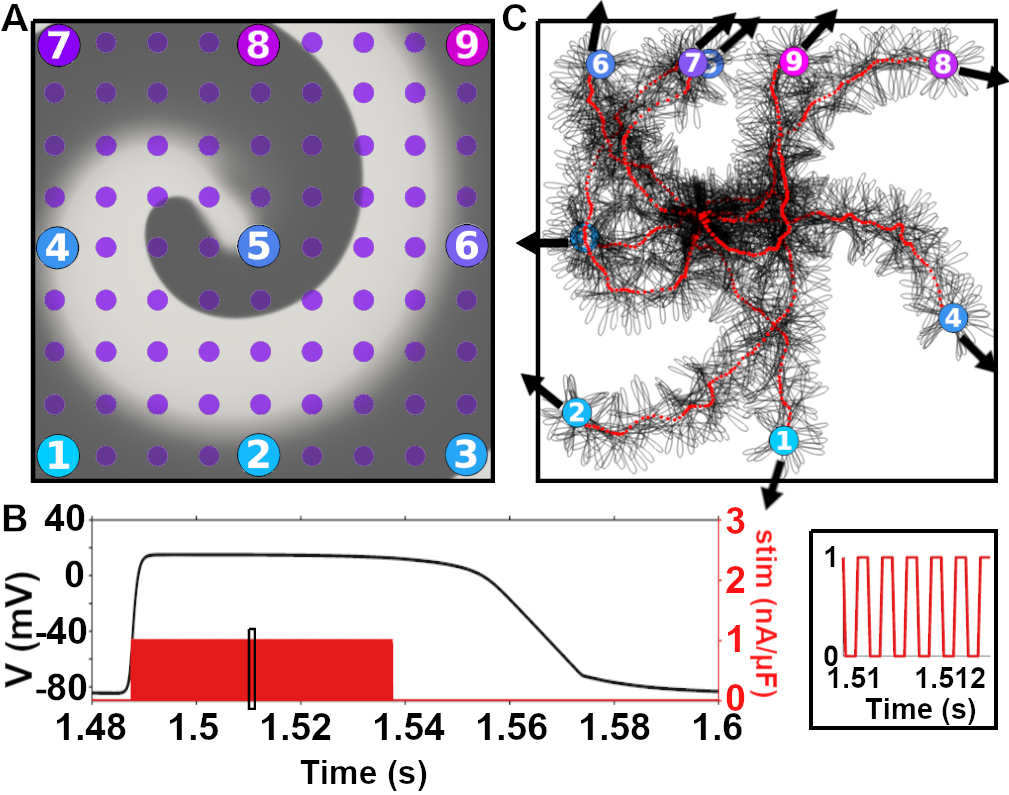} 
	\caption{Adapted RFC enables robust spiral-wave termination. (A) 81 sensor locations used to test robustness of the adapted RFC protocol. (B) Proposed feedback protocol: each sensor trigger initiates a train of 100 square pulses (1 nA/µF, duty cycle 0.5, and period 0.5 ms). Inset shows a zoom-in of the applied pulse train. (C) Resulting COM trajectories for 9 representative sensor locations, labelled on (A). Sustained drift always drives the spiral to the boundary where it is then absorbed.
}
	\label{fig:new} 
\end{figure}
Together, these observations suggest that effective control requires sustained low-amplitude perturbations that drive the spiral through such regions without inducing breakup. We therefore modified RFC by replacing each single feedback pulse with a train of low-amplitude square pulses. This protocol accelerates COM drift while avoiding wave breakup, quasi-stationarity, and sensor-centered confinement, enabling continuous motion toward the boundary and subsequent termination. To test robustness to sensor placement, we performed 81 independent simulations, each with the sensor placed at a different location on a uniform grid spanning the domain (Fig~\ref{fig:new}A). In all cases, the adapted protocol (Fig.~\ref{fig:new}B) produced a sustained COM drift until the spiral reached the boundary and was absorbed (Fig.~\ref{fig:new}C). On the $12\times12$ \unit{\centi\meter\squared} domain, the time required for the spiral to reach the boundary was $22\pm9$ \unit{\second}.

Resonant feedback control has long been proposed as a low-energy strategy for steering reentrant waves toward termination. Classical theory, developed for rigid-core and simpler spirals, relies on resonant forcing to induce directed drift through phase locking~\cite{Davydov1988,Mikhailov1994}. Here, we show that pronounced intrinsic memory fundamentally alters this paradigm. In complex meandering spirals, RFC produces distinct nonlinear regimes---drift, metastable quasi-stationarity, and sensor-centered confinement even without structural anchoring. 
Rather than frequency matching alone, spiral motion reflects the cumulative constructive or destructive interference of phase-dependent perturbations over the meander cycle, which can suppress, localize, or intermittently disrupt drift. Stronger forcing does not overcome these regimes and can instead deform trajectories and trigger wave breakup.

This mechanism also suggests a route to effective control. Rather than increasing stimulus strength, we replace each feedback stimulus with a train of low-amplitude pulses that drives the spiral through quasi-stationary regions while avoiding confinement and wave breakup. The resulting protocol restores sustained directed drift and robust boundary-mediated termination across sensor locations. This is particularly relevant to cardiac systems, where low-energy alternatives to high-energy shocks are actively sought~\cite{Xie97,li2016significance,poole2008prognostic}, including approaches that exploit intrinsic dynamics through low-amplitude, high-frequency, or resonant stimulation~\cite{steinbock1993control,Luther11,hussaini2021drift}. Our results show that, although intrinsic memory constrains conventional RFC, these limitations can be overcome by accounting for the phase-structured response rather than resonance frequency alone. 

\bibliography{apssamp,references_Hans} 

@article{Zaikin1970,
  author  = {Zaikin, A. N. and Zhabotinsky, A. M.},
  title   = {Concentration Wave Propagation in Two-Dimensional Liquid-Phase Self-Oscillating System},
  journal = {Nature},
  volume  = {225},
  number  = {5232},
  pages   = {535--537},
  year    = {1970},
  doi     = {10.1038/225535b0}
}

@article{Winfree1972,
  author  = {Winfree, Arthur T.},
  title   = {Spiral Waves of Chemical Activity},
  journal = {Science},
  volume  = {175},
  number  = {4022},
  pages   = {634--636},
  year    = {1972},
  doi     = {10.1126/science.175.4022.634}
}

@article{Gerisch1987,
  author  = {Gerisch, G.},
  title   = {Cyclic AMP and other signals controlling cell development and differentiation in {Dictyostelium}},
  journal = {Annual Review of Biochemistry},
  year    = {1987},
  volume  = {56},
  pages   = {853--879},
  doi     = {10.1146/annurev.bi.56.070187.004225}
}

@article{Kastberger2008,
  author  = {Kastberger, Gerald and Schmelzer, Erich and Kranner, Ilse},
  title   = {Social waves in giant honeybees repel hornets},
  journal = {PLoS ONE},
  year    = {2008},
  volume  = {3},
  number  = {9},
  pages   = {e3141},
  doi     = {10.1371/journal.pone.0003141}
}

@article{Mantel1996,
  author  = {Mantel, R. M. and Barkley, D.},
  title   = {Periodic forcing of spiral waves in excitable media},
  journal = {Physical Review E},
  volume  = {54},
  number  = {5},
  pages   = {4791--4802},
  year    = {1996},
  doi     = {10.1103/PhysRevE.54.4791}
}

@article{Biktasheva2010,
  author  = {Biktasheva, I. V. and Barkley, D. and Biktashev, V. N. and Foulkes, A. J.},
  title   = {Computation of the drift velocity of spiral waves using response functions},
  journal = {Physical Review E},
  volume  = {81},
  number  = {6},
  pages   = {066202},
  year    = {2010},
  doi     = {10.1103/PhysRevE.81.066202}
}

@article{Davydov1988,
  author  = {Davydov, V. A. and Zykov, V. S. and Mikhailov, A. S. and Brazhnik, P. K.},
  title   = {Drift and Resonance of Helical Waves in Distributed Active Media},
  journal = {Radiophysics and Quantum Electronics},
  volume  = {31},
  number  = {5},
  pages   = {419--426},
  year    = {1988}
}

@article{Mikhailov1994,
  author  = {Mikhailov, A. S. and Davydov, V. A. and Zykov, V. S.},
  title   = {Complex Dynamics of Spiral Waves and Motion of Curves},
  journal = {Physica D: Nonlinear Phenomena},
  volume  = {70},
  number  = {1--2},
  pages   = {1--39},
  year    = {1994},
  doi     = {10.1016/0167-2789(94)90052-3}
}

@article{Courtemanche1998,
  author    = {Courtemanche, Marc and Ramirez, Ricardo J. and Nattel, Stanley},
  title     = {Ionic mechanisms underlying human atrial action potential properties: Insights from a mathematical model},
  journal   = {American Journal of Physiology-Heart and Circulatory Physiology},
  volume    = {275},
  number    = {1},
  pages     = {H301--H321},
  year      = {1998},
  month     = {July},
  doi       = {10.1152/ajpheart.1998.275.1.H301},
  pmid      = {9688927}
}

@article{Grill1995,
  author  = {Grill, S. and Zykov, V. S. and M{\"u}ller, S. C.},
  title   = {Feedback-controlled dynamics of meandering spiral waves},
  journal = {Physical Review Letters},
  year    = {1995},
  volume  = {75},
  number  = {18},
  pages   = {3368--3371},
  doi     = {10.1103/PhysRevLett.75.3368}
}

@article{Cherry08,
    author = {Cherry, EM. and Evans, SJ.},
    title =  {Properties of two human atrial cell models in tissue: restitution, memory, propagation, and reentry}, 
    journal = {J Theor Biol.},
    year = {2008},
    volume = {254},
    number = {3},
    pages = {674-90},
    doi = {10.1016/j.jtbi.2008.06.030.}
}

@article{fenton1998vortex,
  title={Vortex dynamics in three-dimensional continuous myocardium with fiber rotation: Filament instability and fibrillation},
  author={Fenton, Flavio and Karma, Alain},
  journal={Chaos: An Interdisciplinary Journal of Nonlinear Science},
  volume={8},
  number={1},
  pages={20--47},
  year={1998},
  publisher={American Institute of Physics}
}

@article{shajahan2007spiral,
  title={Spiral-wave dynamics depend sensitively on inhomogeneities in mathematical models of ventricular tissue},
  author={Shajahan, TK and Sinha, Sitabhra and Pandit, Rahul},
  journal={Physical Review E—Statistical, Nonlinear, and Soft Matter Physics},
  volume={75},
  number={1},
  pages={011929},
  year={2007},
  publisher={APS}
}

@article{ten2004model,
  title={A model for human ventricular tissue},
  author={Ten Tusscher, Kirsten HWJ and Noble, Denis and Noble, Peter-John and Panfilov, Alexander V},
  journal={American Journal of Physiology-Heart and Circulatory Physiology},
  volume={286},
  number={4},
  pages={H1573--H1589},
  year={2004},
  publisher={American Physiological Society}
}

@article{arndt2022deal,
  title={The deal. II library, version 9.4},
  author={Arndt, Daniel and Bangerth, Wolfgang and Feder, Marco and Fehling, Marc and Gassm{\"o}ller, Rene and Heister, Timo and Heltai, Luca and Kronbichler, Martin and Maier, Matthias and Munch, Peter and others},
  journal={Journal of Numerical Mathematics},
  volume={30},
  number={3},
  pages={231--246},
  year={2022},
  publisher={De Gruyter}
}

@article{roshan2026mechano,
  title={Mechano-electric feedback reduces the occurrence of delayed-afterdepolarization-driven focal activity},
  author={Roshan, Navneet and Pandit, Rahul},
  journal={Computers in Biology and Medicine},
  volume={202},
  pages={111436},
  year={2026},
  publisher={Elsevier}
}

@article{krishnamoorthi2013numerical,
  title={Numerical quadrature and operator splitting in finite element methods for cardiac electrophysiology},
  author={Krishnamoorthi, Shankarjee and Sarkar, Mainak and Klug, William S},
  journal={International journal for numerical methods in biomedical engineering},
  volume={29},
  number={11},
  pages={1243--1266},
  year={2013},
  publisher={Wiley Online Library}
}

@article{wang2013structure,
  title={Structure-based finite strain modelling of the human left ventricle in diastole},
  author={Wang, HM and Gao, H and Luo, XY and Berry, C and Griffith, BE and Ogden, RW and Wang, TJ},
  journal={International journal for numerical methods in biomedical engineering},
  volume={29},
  number={1},
  pages={83--103},
  year={2013},
  publisher={Wiley Online Library}
}

@article{iyer2001experimentalist,
  title={An experimentalist's approach to accurate localization of phase singularities during reentry},
  author={Iyer, Anand N and Gray, Richard A},
  journal={Annals of biomedical engineering},
  volume={29},
  pages={47--59},
  year={2001},
  publisher={Springer}
}

@article{pertsov1993spiral,
  title={Spiral waves of excitation underlie reentrant activity in isolated cardiac muscle.},
  author={Pertsov, Arkady M and Davidenko, Jorge M and Salomonsz, Remy and Baxter, William T and Jalife, Jose},
  journal={Circulation research},
  volume={72},
  number={3},
  pages={631--650},
  year={1993}
}

@article{Rappel2022,
  title = {Intermittent trapping of spiral waves in a cardiac model},
  author = {Rappel, Wouter-Jan},
  journal = {Phys. Rev. E},
  volume = {105},
  issue = {1},
  pages = {014404},
  numpages = {6},
  year = {2022},
  month = {Jan},
  publisher = {American Physical Society},
  doi = {10.1103/PhysRevE.105.014404},
  url = {https://link.aps.org/doi/10.1103/PhysRevE.105.014404}
}

@article{
Xie97,
author = {Jianlin Xie  and Max Harry Weil  and Shijie Sun  and Wanchun Tang  and Yoji Sato  and Xiaohua Jin  and Joe Bisera },
title = {High-Energy Defibrillation Increases the Severity of Postresuscitation Myocardial Dysfunction},
journal = {Circulation},
volume = {96},
number = {2},
pages = {683-688},
year = {1997}
}

@article{li2016significance,
  title={The significance of shocks in implantable cardioverter defibrillator recipients},
  author={Li, Anthony and Kaura, Amit and Sunderland, Nicholas and Dhillon, Paramdeep S and Scott, Paul A},
  journal={Arrhythmia \& electrophysiology review},
  volume={5},
  number={2},
  pages={110},
  year={2016}
}

@article{poole2008prognostic,
  title={Prognostic importance of defibrillator shocks in patients with heart failure},
  author={Poole, Jeanne E and Johnson, George W and Hellkamp, Anne S and Anderson, Jill and Callans, David J and Raitt, Merritt H and Reddy, Ramakota K and Marchlinski, Francis E and Yee, Raymond and Guarnieri, Thomas and others},
  journal={New England Journal of Medicine},
  volume={359},
  number={10},
  pages={1009--1017},
  year={2008},
  publisher={Mass Medical Soc}
}

@article {Luther11,
	Title = {Low-energy control of electrical turbulence in the heart},
	Author = {Luther, Stefan and Fenton, Flavio H and Kornreich, Bruce G and Squires, Amgad and Bittihn, Philip and Hornung, Daniel and Zabel, Markus and Flanders, James and Gladuli, Andrea and Campoy, Luis and Cherry, Elizabeth M and Luther, Gisa and Hasenfuss, Gerd and Krinsky, Valentin I and Pumir, Alain and Gilmour, Robert F and Bodenschatz, Eberhard},
	DOI = {10.1038/nature10216},
	Number = {7355},
	Volume = {475},
	Month = {July},
	Year = {2011},
	Journal = {Nature},
	ISSN = {0028-0836},
	Pages = {235—239},
	URL = {https://europepmc.org/articles/PMC3153959},
}

@article{hussaini2021drift,
  title={Drift and termination of spiral waves in optogenetically modified cardiac tissue at sub-threshold illumination},
  author={Hussaini, Sayedeh and Venkatesan, Vishalini and Biasci, Valentina and Romero Sep{\'u}lveda, Jos{\'e} M and Qui{\~n}onez Uribe, Raul A and Sacconi, Leonardo and Bub, Gil and Richter, Claudia and Krinski, Valentin and Parlitz, Ulrich and others},
  journal={Elife},
  volume={10},
  pages={e59954},
  year={2021},
  publisher={eLife Sciences Publications, Ltd}
}

@article{toal2009short,
  title={Short-term memory and restitution during ventricular fibrillation in human hearts: an in vivo study},
  author={Toal, Satish C and Farid, Talha A and Selvaraj, Raja and Chauhan, Vijay S and Masse, Stephane and Ivanov, Joan and Harris, Louise and Downar, Eugene and Franz, Michael R and Nanthakumar, Kumaraswamy},
  journal={Circulation: Arrhythmia and Electrophysiology},
  volume={2},
  number={5},
  pages={562--570},
  year={2009},
  publisher={Lippincott Williams \& Wilkins}
}

@article{jordan2004determining,
  title={Determining the effects of memory and action potential duration alternans on cardiac restitution using a constant-memory restitution protocol},
  author={Jordan, Peter N and Christini, David J},
  journal={Physiological measurement},
  volume={25},
  number={4},
  pages={1013--1024},
  year={2004}
}

@article{weiss2006pulsus,
  title={From pulsus to pulseless: the saga of cardiac alternans},
  author={Weiss, James N and Karma, Alain and Shiferaw, Yohannes and Chen, Peng-Sheng and Garfinkel, Alan and Qu, Zhilin},
  journal={Circulation research},
  volume={98},
  number={10},
  pages={1244--1253},
  year={2006},
  publisher={Lippincott Williams \& Wilkins}
}

@article{decker2009properties,
  title={Properties and ionic mechanisms of action potential adaptation, restitution, and accommodation in canine epicardium},
  author={Decker, Keith F and Heijman, Jordi and Silva, Jonathan R and Hund, Thomas J and Rudy, Yoram},
  journal={American Journal of Physiology-Heart and Circulatory Physiology},
  volume={296},
  number={4},
  pages={H1017--H1026},
  year={2009},
  publisher={American Physiological Society}
}

@article{steinbock1993control,
  title={Control of spiral-wave dynamics in active media by periodic modulation of excitability},
  author={Steinbock, Oliver and Zykov, Vladimir and M{\"u}ller, Stefan C},
  journal={Nature},
  volume={366},
  number={6453},
  pages={322--324},
  year={1993},
  publisher={Nature Publishing Group UK London}
}

@article{luo1991model,
  title={A model of the ventricular cardiac action potential. Depolarization, repolarization, and their interaction.},
  author={Luo, Ching-hsing and Rudy, Yoram},
  journal={Circulation research},
  volume={68},
  number={6},
  pages={1501--1526},
  year={1991}
}

@article{Braune2000,
  author  = {Braune, M. and Engel, H.},
  title   = {Feedback-controlled forcing of meandering spiral waves in an open gel reactor},
  journal = {Physical Review E},
  volume  = {62},
  number  = {5},
  pages   = {5986--5988},
  year    = {2000},
  doi     = {10.1103/PhysRevE.62.5986}
}

@article{Zykov2002,
  author  = {Zykov, Vladimir S. and Engel, Harald},
  title   = {Resonance attractors of spiral waves in excitable media under global feedback},
  journal = {Physical Review E},
  volume  = {66},
  number  = {1},
  pages   = {016206},
  year    = {2002},
  doi     = {10.1103/PhysRevE.66.016206}
}

@article{Zykov2004,
  author  = {Zykov, V. S. and Bordiougov, G. and Brandtst{\"a}dter, H. and Gerdes, I. and Engel, H.},
  title   = {Global Control of Spiral Wave Dynamics in an Excitable Domain of Circular and Elliptical Shape},
  journal = {Physical Review Letters},
  volume  = {92},
  number  = {1},
  pages   = {018304},
  year    = {2004},
  doi     = {10.1103/PhysRevLett.92.018304}
}

@ARTICLE{Biktashev:1995,
        AUTHOR = "Biktashev, V.N.  and Holden, A.V. ",
        TITLE = "Control of re-entrant activity in a model of mammalian atrial tissue",
        JOURNAL = "Proc. Roy. Soc. Lond. B",
        VOLUME  = "260",
        PAGES   = "211-217",
        YEAR    = "1995"
      }

@ARTICLE{Biktashev:1995b,
        AUTHOR = "Biktashev, V.N.  and Holden, A.V. ",
        TITLE = "Resonant drift of autowave vortices in two dimensions and the effect of boundaries and inhomogeneities",
        JOURNAL = "Chaos,  Solitons and Fractals",
        VOLUME  = "5",
        PAGES   = "575-562",
        YEAR    = "1995"
}

@article{Dierckx:2017PRL,
  title = {Filament Tension and Phase-Locking of Meandering Scroll Waves},
  author = {Dierckx, H. and Verschelde, H. and Panfilov, A.V. and Biktasheva, I.V. and Biktashev, V.N.},
  journal = {Phys. Rev. Lett.},
  volume = {119},
pages={258101 1-6},
  year = {2017}
}

@ARTICLE{Zykov:1997,
        AUTHOR = "Zykov, V.S.  and Mikhailov, A.S.  and M{\"u}ller,  S.C. ",
        TITLE = "Controlling spiral waves in confined geometries by global feedback",
        JOURNAL = "Phys. Rev. Lett.  ",
        VOLUME  = "78",
        PAGES   = "3398-3401",
        YEAR    = "1997"   }
\cleardoublepage

\appendix

\section*{End Matter}

\subsection*{Cardiac excitation model}

Simulations were performed using the three-variable Fenton--Karma (3vFK) model of cardiac excitation~\cite{fenton1998vortex}. In this formulation, the dimensionless transmembrane voltage \(u \in [0,1]\) evolves according to the reaction–diffusion equation
\begin{equation}
\partial_{t} u
=
\nabla \cdot \left( D_0\nabla u \right)
-
\mathcal{J}_{\text{ion}},
\label{eq:rxn_diff_pde1}
\end{equation}
where electrical coupling is represented by an isotropic diffusion constant \(D_0 = 0.001~\mathrm{cm^2\,ms^{-1}}\)~\cite{shajahan2007spiral}, which yields a plane-wave conduction velocity of approximately \(67~\mathrm{cm\,s^{-1}}\), consistent with measurements in human ventricular tissue~\cite{ten2004model}.
The total ionic current is decomposed as
\[
\mathcal{J}_{\text{ion}} = J_\mathrm{fi} + J_\mathrm{so} + J_\mathrm{si},
\]
where
\begin{align}
J_{\mathrm{fi}}(u; v) &=
-\frac{v}{\tau_\mathrm{d}} \Theta\left(u-u_\mathrm{c}\right)(1-u)\left(u-u_\mathrm{c}\right),\\
J_{\mathrm{so}}(u) &=
\frac{u}{\tau_\mathrm{o}} \Theta\left(u_\mathrm{c}-u\right)
+
\frac{1}{\tau_\mathrm{r}} \Theta\left(u-u_\mathrm{c}\right),\\
J_{\mathrm{si}}(u; w) &=
-\frac{w}{2\tau_{\mathrm{si}}}
\left(1+\tanh\left[k\left(u-u_\mathrm{c}^{\mathrm{si}}\right)\right]\right).
\end{align}
Fast and slow recovery dynamics are governed by gating variables \(v\) and \(w\):
\begin{align}
\partial_{t} v &=
\frac{\Theta\left(u_{c}-u\right)(1-v)}{\tau_{v}^{-}(u)}
-
\frac{\Theta\left(u-u_{c}\right)v}{\tau_{v}^{+}},\\
\partial_{t} w &=
\frac{\Theta\left(u_{c}-u\right)(1-w)}{\tau_{w}^{-}}
-
\frac{\Theta\left(u-u_{c}\right)w}{\tau_{w}^{+}}.
\end{align}
Model parameters were selected to produce a five-petal complex meandering spiral. Unless otherwise noted,\(
u_c = 0.13,\;
\tau_v^{+} = 10~\mathrm{ms},\;
\tau_w^{-} = 65~\mathrm{ms},\;
\tau_w^{+} = 1000~\mathrm{ms},\;
\tau_d = 1/8.7~\mathrm{ms},\;
\tau_0 = 12.5~\mathrm{ms},\;
\tau_r = 25~\mathrm{ms},\;
\tau_{si} = 22.22~\mathrm{ms}.
\)
Dimensional voltage was obtained via \(V = 100u - 85\) (mV), yielding a single-cell action potential duration of approximately 210 ms at 90\% repolarization.

\subsection*{Numerical methods}

Equation~\ref{eq:rxn_diff_pde1} was solved either on planar two-dimensional domains or on a two-dimensional surface embedded in three-dimensional space. Unless stated 2D simulations were performed on a \(8 \times 8~\mathrm{cm^2}\) domain using finite-difference method discretized with \(0.02 \times 0.02~\mathrm{cm^2}\) elements and time step \(\delta t = 0.05~\mathrm{ms}\). Reaction kinetics were integrated locally at mesh nodes using the Rush--Larsen scheme~\cite{krishnamoorthi2013numerical,roshan2026mechano}.

Zero-flux (Neumann) boundary conditions were imposed on the diffusion operator for all domain boundaries in both two- and three-dimensional simulations. Time-step convergence and mesh-refinement studies confirmed numerical stability and spatial resolution adequacy.

\subsection*{Anatomical surface simulations}
For anatomically realistic studies, dynamics were restricted to the epicardial surface of the human left ventricle, modeled as a co-dimension-one manifold \(\Gamma \subset \mathbb{R}^3\). The governing equation becomes
\begin{equation}
\label{eq:PDE2}
\partial_{t} u
=
D_0 \nabla_\Gamma \cdot \nabla_\Gamma u
-
\mathcal{J}_{\text{ion}},
\end{equation}
where \(\nabla_\Gamma\) denotes the surface gradient. The ventricular geometry was discretized into quadrilateral elements of diameter \(0.021\text{--}0.044~\mathrm{cm}\)~\cite{wang2013structure}.

Equation~\ref{eq:PDE2} was solved using second-order Strang operator splitting. The diffusion term was integrated with a semi-implicit Crank--Nicolson scheme implemented in the finite-element library deal.II~\cite{arndt2022deal} using bilinear quadrilateral elements. Reaction kinetics were integrated locally at mesh nodes using the Rush--Larsen scheme.

Spiral waves were always initiated using the standard S1--S2 cross-field protocol~\cite{pertsov1993spiral}, and tip trajectories were identified using the phase-singularity method of Iyer and Gray~\cite{iyer2001experimentalist}.

\subsection*{Modeling heterogeneity}

Spatial ionic heterogeneity was introduced by independently perturbing key recovery time constants of the 3vFK model. Specifically, \(\tau(\mathbf{x})=\tau_0\left[1+\epsilon(\mathbf{x})\right],\)
where \(\epsilon(\mathbf{x}) \sim \mathcal{U}[-0.1,\,0.1]\), corresponding to \(\pm10\%\) variability. Parameters perturbed included
\(\tau_{v1}^{-}, \tau_{v2}^{-}, \tau_v^{+}, \tau_0, \tau_r, \tau_{si}, \tau_w^{-}, \tau_w^{+}\).
Random fields were generated with reproducible seeds and were spatially uncorrelated at the grid scale. Diffusion remained spatially uniform.

Structural conduction heterogeneity was introduced via randomly distributed electrically uncoupled sites. A fixed fraction of grid points (typically 3\%) was designated as nonconducting using a pseudorandom mask. At these locations, local ionic dynamics were preserved but diffusive coupling to neighboring nodes was eliminated by enforcing zero-flux conditions across obstacle interfaces.
Sensor and stimulus locations were excluded from obstacle placement. Unless otherwise noted, obstacle realizations were spatially uncorrelated.

\subsection*{Resonance-based feedback protocol (RFC)}

RFC was implemented by designating a single grid location as the sensor. The local voltage \(u\) at this site was continuously monitored, and a spatially uniform stimulus was delivered whenever the signal crossed a prescribed activation threshold. Stimulus amplitude was held constant within each simulation and varied between \(6\) and \(20~\mathrm{nA/\mu F}\) across simulations. Unless otherwise stated, the stimulus duration was \(0.5~\mathrm{ms}\). For schematic representation, see Fig.~\ref{fig:Fig1}C.

\subsection*{Robustness of dynamical regimes}
\begin{figure}[h!] 
	\centering
	\includegraphics[width=8.54cm]{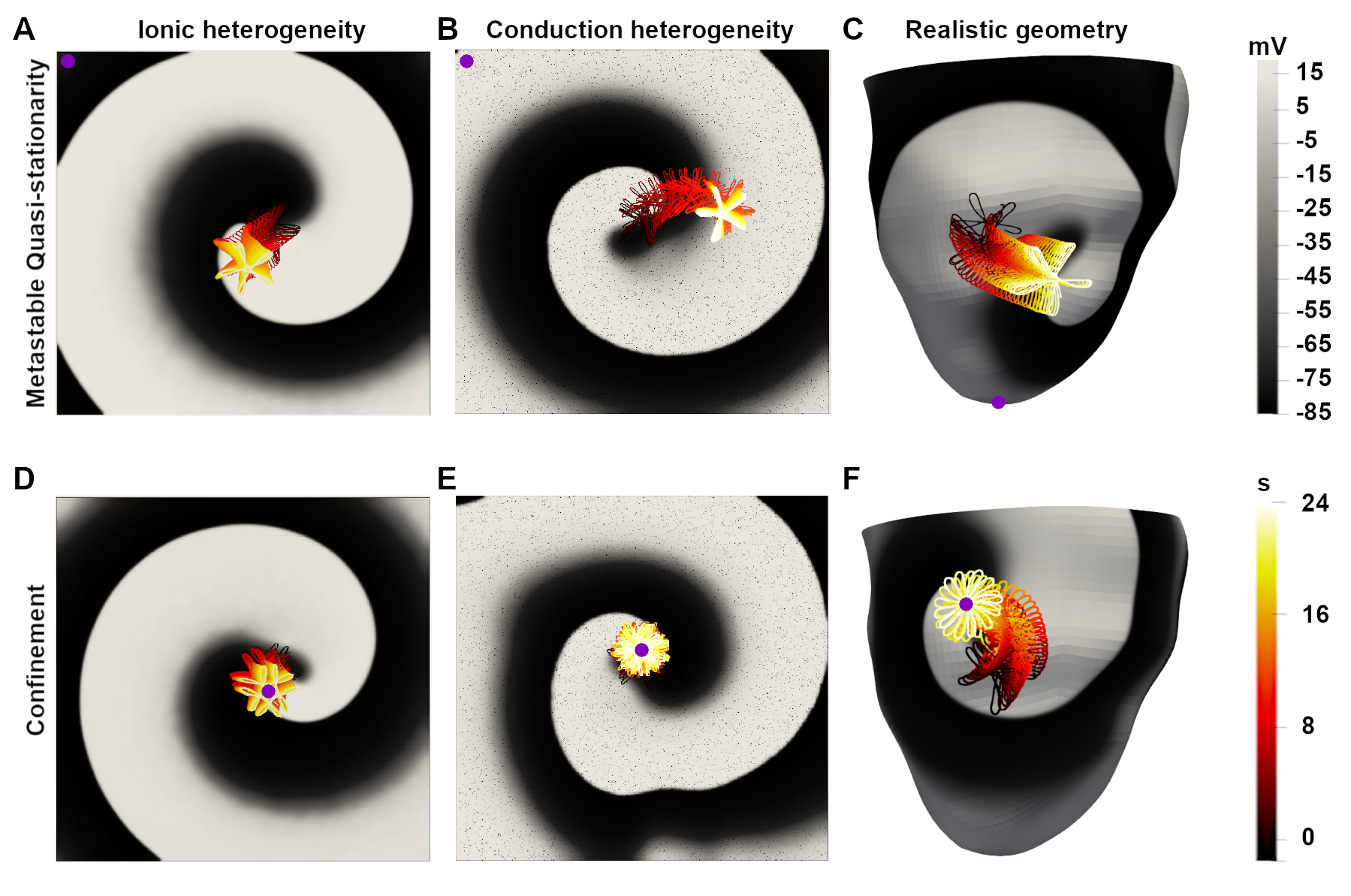} 
	\caption{Robustness of quasi-stationary and confinement regimes under physiological complexity.
(A,D) Ionic heterogeneity (±10\%): both regimes persist, exhibiting ultra-slow drift or stable confinement over $>20$~s.
(B,E) Conduction inhomogeneity (3\% obstacle fraction): dynamics remain organized into quasi-stationary motion or confinement.
(C,F) Anatomically realistic geometry: both regimes persist on the human left-ventricular epicardial surface despite curvature and structural complexity.
Sensor locations are marked by filled purple circles.}
	\label{fig:Supp1} 
\end{figure}

We tested the persistence of quasi-stationary and confinement regimes under physiological complexity by introducing ionic heterogeneity, conduction inhomogeneity, and realistic ventricular geometry. In all cases, both regimes persisted over $>20$~s of RFC (Fig.~\ref{fig:Supp1} A-F). These results establish both regimes as robust features of resonantly paced spiral waves in cardiac tissue.

\end{document}